# Monitoring of optical properties of deep waters of Lake Baikal in 2021-2022

**V.M. Aynutdinov[a], V.A. Allakhverdyan[b], A.D. Avrorin[a], A.V. Avrorin[a], Z. Bardačová[c,d], I.A. Belolaptikov[b], E.A. Bondarev[a], I.V. Borina[b], N.M. Budnev[e], V.A. Chadymov[l], A.S. Chepurnov[f], V.Y. Dik[b,g], G.V. Domogatsky[a], A.A. Doroshenko[a], R. Dvornický[c,*], A.N. Dyachok[e], Zh.-A.M. Dzhilkibaev[a], E. Eckerová[c,d], T.V. Elzhov[b], L. Fajt[d], V.N. Fomin[l], A.R. Gafarov[e], K.V. Golubkov[a], N.S. Gorshkov[b], T. I. Gress[e], K.G. Kebkal[h], I.V. Kharuk[a], E.V. Khramov[b], M.M. Kolbin[b], S.O. Koligaev[i], K.V. Konischev[b], A.V. Korobchenko[b], A.P. Koshechkin[a], V.A. Kozhin[f], M.V. Kruglov[b], V.F. Kulepov[j], Y.E. Lemeshev[e], M.B. Milenin[a,†], R.R. Mirgazov[e], D.V. Naumov[b], A.S. Nikolaev[f], D.P. Petukhov[a], E.N. Pliskovsky[b], M.I. Rozanov[k], E.V. Ryabov[e], G.B. Safronov[a], D. Seitova[b,g], B.A. Shaybonov[b], M.D. Shelepov[a], S.D. Shilkin[a], E.V. Shirokov[f], F. Šimkovic[c,d], A.E. Sirenko[b], A.V. Skurikhin[f], A.G. Solovjev[b], M.N. Sorokovikov[b], I. Štekl[d], A.P. Stromakov[a], O.V. Suvorova[a], V.A. Tabolenko[e], B.B. Ulzutuev[b], Y.V. Yablokova[b], D.N. Zaborov[a], S.I. Zavyalov[b], and D.Y. Zvezdov[b]**

[a] *Institute for Nuclear Research, Russian Academy of Sciences, Moscow, Russia*

[b] *Joint Institute for Nuclear Research, Dubna, Russia*

[c] *Comenius University, Bratislava, Slovakia*

[d] *Czech Technical University in Prague, Institute of Experimental and Applied Physics, Czech Republic*

[e] *Irkutsk State University, Irkutsk, Russia*

[f] *Skobeltsyn Institute of Nuclear Physics, Moscow State University, Moscow, Russia*

[g] *Institute of Nuclear Physics of the Ministry of Energy of the Republic of Kazakhstan, Almaty, Kazakhstan*

[h] *LATENA, St. Petersburg, Russia*

[i] *INFRAD, Dubna, Russia*

[j] *Nizhny Novgorod State Technical University, Nizhny Novgorod, Russia*

[k] *St. Petersburg State Marine Technical University, St. Petersburg, Russia*

[l] *Moscow, free researcher*

E-mail: e_v_ryabov@mail.ru

*Speaker

†Deceased

*We present the results of the two-year (2021-2022) monitoring of absorption and scattering lengths of light with wavelength 400–620 nm within the effective volume of the deep underwater neutrino telescope Baikal-GVD, which were measured by a device «BAIKAL-5D» №2. The «BAIKAL-5D» №2 was installed during the 2021 winter expedition at a depth of 1180 m. The absorption and scattering lengths were measured every week in 9 spectral points. The device «BAIKAL-5D» №2 also has the ability to measure detailed scattering and absorption spectra. The data obtained make it possible to estimate the range of changes in the absorption and scattering lengths over a sufficiently long period of time and to investigate the relationship between the processes of changes in absorption and scattering. An analysis was made of changes in absorption and scattering spectra for the period 2021-2022.*

## 1. Introduction

The underwater neutrino telescope Baikal-GVD uses photomultiplier tubes (PMTs) for collecting the Cherenkov light emitted by the charged particles passing through the effective volume of the detector. The array PMTs are installed in the deep water of lake Baikal. Therefore, knowledge of the optical properties of the deep water of lake Baikal is essential for the operation of Baikal-GVD as well as for the analysis of the collected data. The scattering and absorption of light is quantitatively described by the so-called hydro-optical characteristics, i.e. absorption length $L_a(\lambda)$ or absorption coefficient $a(\lambda) = 1/L_a(\lambda)$, whereas for the scattering of light, the relevant quantities are the scattering length $L_b(\lambda)$, or scattering coefficient $b(\lambda) = 1/L_b(\lambda)$, and scattering function $\chi(\gamma, \lambda)$, with $\lambda$ being the wavelength and $\gamma$ as the scattering angle [1].

For in situ long-term measurements of the scattering and absorption of light, the staff of the Institute of Applied Physics ISU developed a series of devices «BAIKAL-5D». During the winter expeditions of 2020 and 2021, two instruments, «BAIKAL-5D» №1 and «BAIKAL-5D» №2, were installed in the deployment area of the underwater neutrino telescope Baikal-GVD at depths of 1250 m and 1180 m, respectively.

## 2. Device structure and measurement method

The device (Fig. 1, b) has a hermetically sealed main housing (labeled "1" on the figure), a point-like quasi-isotropic emitter (2), a light receiver displacement drive (3), a direct light source shading system (a screen and a stepper motor , labeled 4 and 5, respectively) and the wide-angle receiver (6) in a separate housing which is moved on thin cables so that the distance between the source and the receiver can vary in increments of 0.64 mm within 1.5-7.5 m. In more detail, the layout of the «BAIKAL-5D» instrument and the principle of operation are described in [2].

The methodological basis for measuring the absorption and scattering by a device like

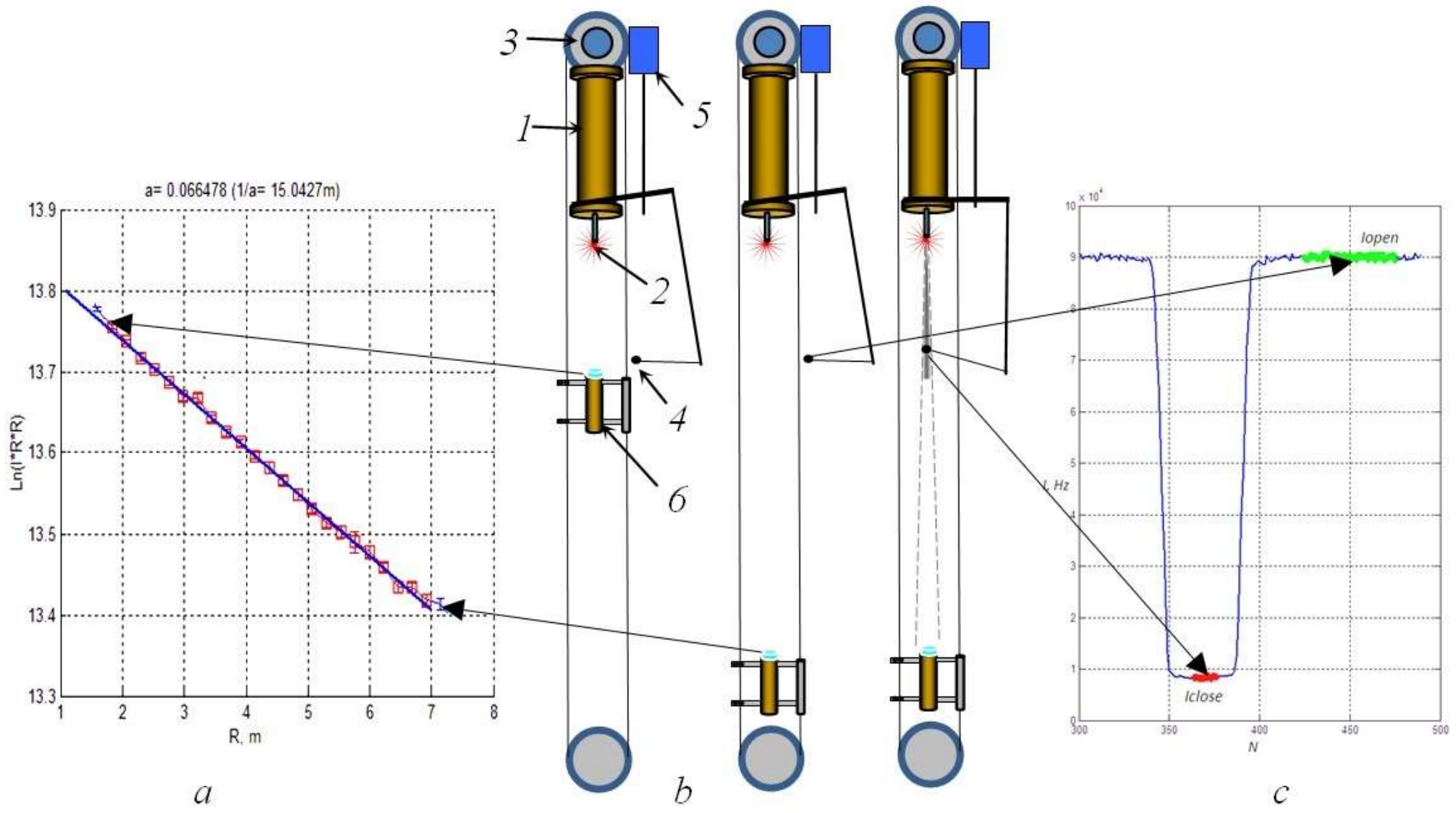

«BAIKAL-5D» is discussed in detail in [3, 4, 5]. It should be noted that the design of the «BAIKAL-5D» instruments allows only the forward scattering component (excluding scattering at an angle of less than 1°) to be measured. For deep lake Baikal water, the estimation of *b* coinsides with forward scattering coefficient $b_{fwd}$ within experemental error [4]. The algorithm used in this study for measuring the dependencies of absorption and scattering of light is as follows.

**Figure 1**. Layout of the «BAIKAL-5D» instrument with movable parts shown in three different positions (b), algorithm of measurement of absorption length (a) and scattering length (c).

Measurement of absorption. Algorithm A1: for a fixed wavelength, the dependence of the average intensity value $I_k$ on the distance $R_k$ between the receiver and the source is measured:

$$I_k = \frac{1}{m}\sum_{j=1}^{m} I_j(R_k), \quad R_k = R_{\min} + \frac{R_{\max} - R_{\min}}{n} k, \quad k \in [0, n], \quad n \in [16,32],$$

where *n* is the total number of distance points, and *m* is the number of measurements in each point (Fig. 1, a). A linear approximation of the sequence of values $\{R_k, \ln(I_k R_k^2)\}$:

$$\ln(I_k R_k^2) = -aR_k + const, \quad a = 1/L_a$$

is performed using the method of least squares in the distance range of 1.5 to 7.0 meters.

Algorithm A2: For two distances, $R_1$ and $R_2$, the spectra of the primary source, $I_1(\lambda)$ and $I_2(\lambda)$, are measured in the wavelength range of 390-620nm. The spectral dependence of the absorption coefficient is calculated by the formula:

$$a(\lambda) = \left[\ln(I_1(\lambda)R_1^2) - \ln(I_2(\lambda)R_2^2)\right]/(R_2 - R_1).$$

Measurement of scattering. Algorithm S1: for the fixed wavelength and maximum distance *R*=7.5 m from the source to the receiver, time-averaged intensity values $I_{open}$ and $I_{close}$ are measured with an open and shaded source (Fig. 1, c). The scattering coefficient is calculated by the formula:

$$b = 1/L_b = -\ln(1 - I_{close}/I_{open})/R.$$

Algorithm S2. This algorithm uses the measurement of the spectra of the primary source in the range of 390-620 nm with an open $I_{open}(\lambda)$ and shaded $I_{close}(\lambda)$ source. The spectral dependence of the scattering coefficient is calculated by the formula:

$$b(\lambda) = 1/L_b(\lambda) = -\ln(1 - I_{close}(\lambda)/I_{open}(\lambda))/R.$$

Algorithms A1 and S1, due to data averaging and shorter measurement time, make it possible to obtain the values of absorption and scattering coefficients with a smaller error than algorithms A2 and S2, in which data averaging is problematic due to the factor of temporal instability of the light source and the measuring channel.

## 3. Measurement results

During the period 2021-2022, the «BAIKAL-5D» №2 (depth 1180 m) device worked all the time, excluding the winter expedition of 2022, and the «BAIKAL-5D» №1 (depth 1250 m) device - only from April 2022. Below we will present the results of the measurement with the «BAIKAL-5D» №2, using the data from the «BAIKAL-5D» №1 for comparison. The random component of measurement error of the absorption coefficient, using the A1 algorithm, was 2-4% and of the forward scattering coefficient (excluding scattering at an angle of less than 1°), using the S1 algorithm, was 3-7%.

### 3.1 The time dependence of absorption and scattering

The absorption and scattering were measured (using algorithms A1 and S1) regularly once a week at 9 wavelengths (405nm, 430nm, 450nm, 460nm, 490nm, 510nm, 532nm, 560nm, 590nm) and at 3 wavelengths (405nm, 460nm, 532nm), respectively. The absorption spectrum (algorithm A2) was also measured once a week, and the scattering spectrum (algorithm S2) was measured approximately every month.

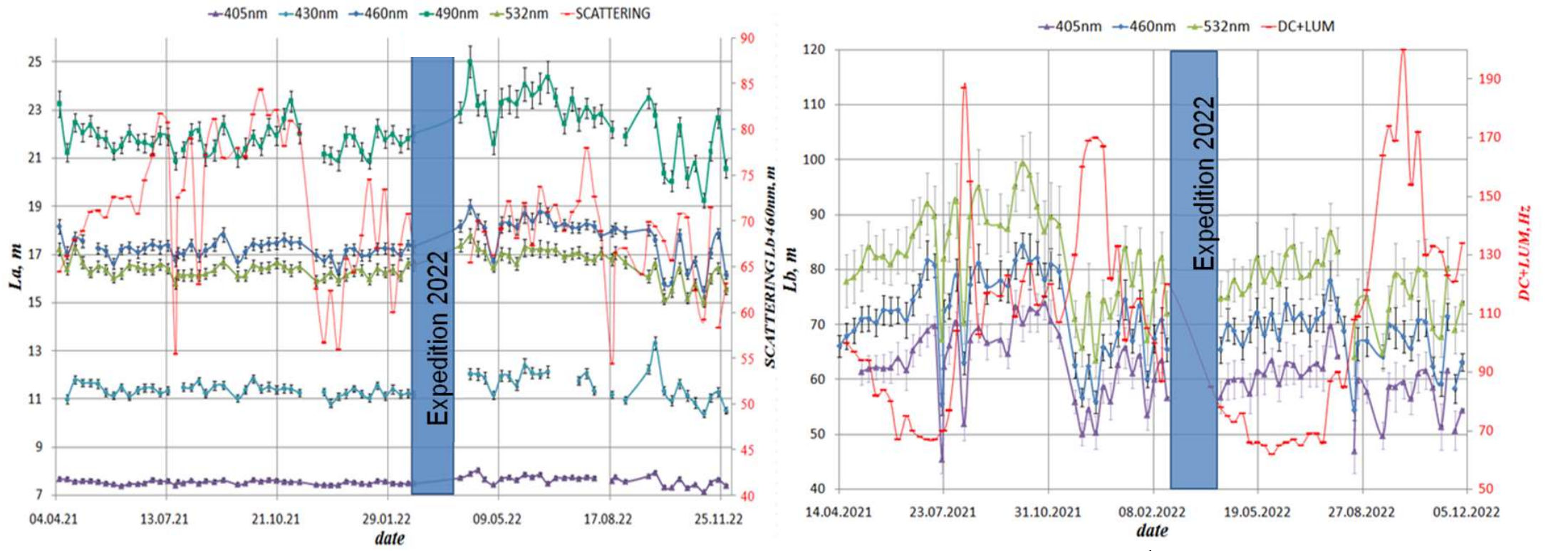

a. b.

**Figure 2.** The time dependence of absorption (a) and scattering (b) lengths in 2021-2022. On the right axes of the graphs are plotted to compare the time dependence of scattering length for λ=460nm (Fig. 2,a) and the time dependence of total signal of the dark current and the glow of water (Fig. 2, b).

The time dependence of absorption and scattering lengths are shown in Fig.2. Using the long-term monitoring data (period 2021-2022, depth 1180 m) we find variations of the absorption coefficient to be within a ±13% interval and variations of the scattering coefficient within ±25%, which are slightly less than observed in 2020 at a depth of 1250m, when an increase in water transparency was observed in the second half of the year [2]. There is a good correlation between the change in absorption for different wavelengths of the spectrum (this fact also holds for the scattering). For example, the correlation coefficient between time dependences of $L_a(\lambda = 460\text{nm})$ and $L_a(\lambda = 532\text{nm})$ is 0.92, and the correlation coefficient between time dependences of $L_b(\lambda = 460\text{nm})$ and $L_b(\lambda = 532\text{nm})$ is 0.94.

The measurements of the scattering length show evidence of significant short-term minima (Fig. 2, b), usually in the summer months; similar effects were also observed in 2020 (at a depth of 1250 m). The variations in the luminescence of water, estimated from the recorded total PMT count rate (dark current rate plus the luminescence of water), appear to anti-correlate with the variations of the scattering length (Fig. 2, b).

### 3.2 Absorption and scattering spectra

The absorption spectral dependence of deep waters of Lake Baikal, obtained using algorithms A1 and A2, is shown in Fig 3, a. All absorption spectra measured at other times have a very similar shape, with a synchronous change in absorption with time at all wavelengths within ±15%. The registered inflection of the spectrum at 600 nm is associated with the absorption of pure water [6] and confirms the correctness of our measurements. There is a good agreement between the data obtained using algorithms A1 and A2.

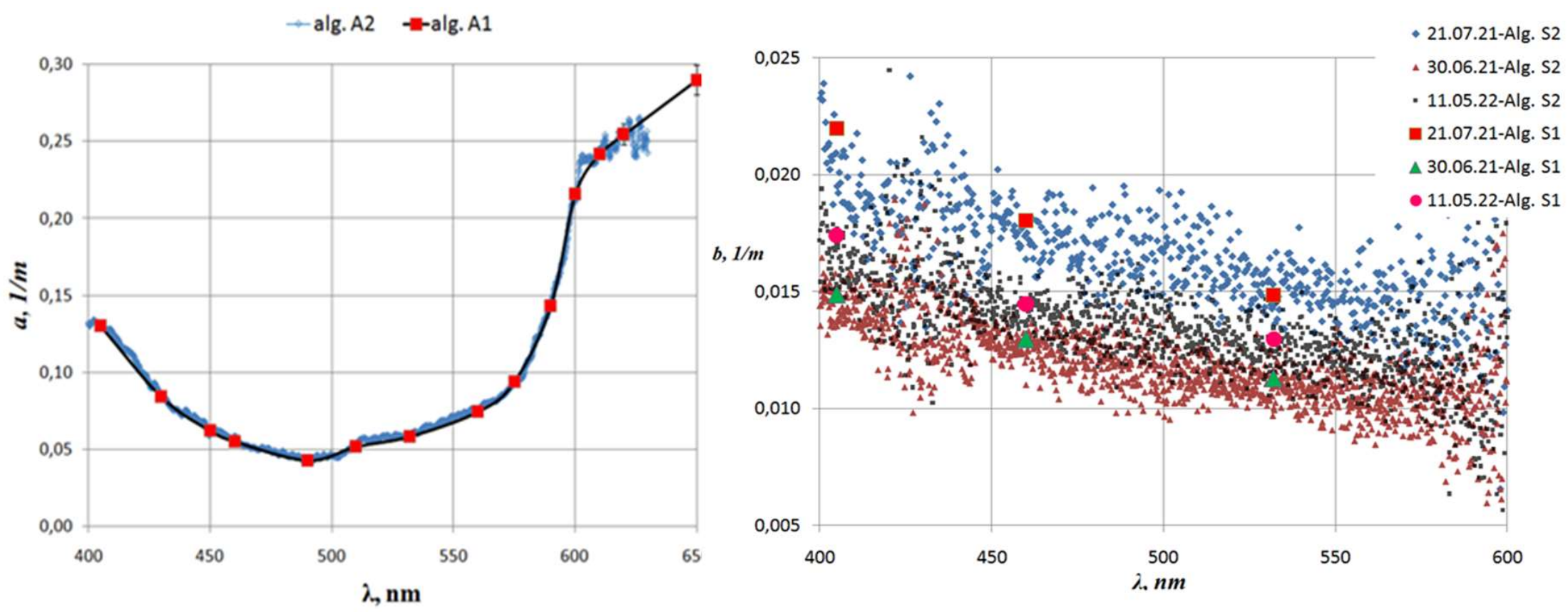

**Figure 3.** Absorption spectrum 26.04.2022, depth 1180m (a) and scattering spectra (b).

The scattering spectra obtained using the S2 algorithm (Fig. 3, b) have a significant random measurement error (~15 %) associated with the low intensity of the recorded light with a shaded source and the lack of averaging of the measured signal at each point of the spectrum.

The more accurate measurements of the scattering coefficient obtained using the S1 algorithm are characterized by a random measurement error of less than 6% and are consistent, within the measurement errors, with the measurements obtained with algorithm S2. The shape of the scattering spectral dependence is found to be approximately hyperbolic at wavelengths of 400-600 nm.

## 4. Conclusion

In 2021-2022, we managed to implement continuous *in situ* monitoring of absorption and scattering lengths of light with wavelength of 400–620 nm within the effective volume of the deep underwater neutrino telescope Baikal-GVD. Using the long-term monitoring data (period 2021-2022, depth 1180 m) we find variations of the absorption coefficient to be within a ±13% interval and variations of the scattering coefficient within ±25%. The monitoring of the absorption spectral dependence allowed us to conclude that its shape is stable over time within the measurement uncertainties (~ 9%), at least at the depth of 1180 m. The values of the absorption and scattering coefficients coincide with good accuracy with the previously obtained data [7] and the 2020 measurements by «BAIKAL-5D» №1.

## 5. Acknowledgments

On behalf of the author, gratitude is expressed to the staff of Institute of Applied Physics ISU, as well as to all participants of the *Baikal-GVD Collaboration* for the provided data and technical support for measurements of hydrooptical characteristics during 2021-2022.